\documentclass{ar-1col}

\usepackage[comma]{natbib}
\usepackage{enumerate}
\usepackage{amssymb}                 
\usepackage{amsfonts}
\usepackage{amsmath}
\usepackage{array,color}
\usepackage{xcolor}
\usepackage{cases}
\usepackage{mathrsfs}
\usepackage{graphicx,epsfig}
\usepackage{float}
\usepackage{subfigure}
\usepackage{diagbox}
\usepackage{mathtools}
\usepackage{multicol,times}

\renewcommand{\theequation}{(\arabic{equation})}
\usepackage{float}

\newcommand{\pr}{{\rm Pr}}

\newcommand{\bq}{\begin{quote}}
\newcommand{\eq}{\end{quote}}
\newcommand{\beq}{\begin{equation}}
\newcommand{\eeq}{\end{equation}}
\newcommand{\beqn}{\begin{eqnarray}}
\newcommand{\eeqn}{\end{eqnarray}}

\newcommand{\CN}{{\cal N}}

\newcommand{\RR}{{\bf R}}

\newcommand{\bfW}{{\mathbf W}}
\newcommand{\hsp}{{\hspace{2ex}}}
\newcommand{\by}{{\mathbf y}}
\newcommand{\bfq}{{\mathbf q}}
\newcommand{\bu}{{\mathbf u}}

\newcommand{\bw}{{\mathbf w}}
\newcommand{\bth}{{\mathbf \theta}}

\newcommand{\calZ}{{\mathcal Z}}
\newcommand{\calX}{{\mathcal X}}

\newcommand{\eps}{{\epsilon}}

\newcommand{\sil}{\mbox{SL}}

\def\ld{\ldots}

\def\X{{\cal X}}

\def\hbar{\overline{h}}

\def\one{{\bf 1}}

\begin{document}

\markboth{Craiu and Levi}{Approximate Bayesian Computation}

\title{Approximate Methods for Bayesian Computation}

\author{Radu V. Craiu,$^1$ and Evgeny Levi,$^2$ 
\affil{$^1$Department of Statistical Sciences, University of Toronto, Toronto, Canada, M5G 1Z5; email: radu.craiu@utoronto.ca}
\affil{$^2$Department of Statistical Sciences, University of Toronto, Toronto, Canada, M5G 1Z5}}


\begin{abstract}

Rich data generating mechanisms are ubiquitous in this age of information and require complex statistical models to draw meaningful inference. While Bayesian analysis has seen enormous development in the last 30 years, benefitting from the impetus given by the successful application of Markov chain Monte Carlo (MCMC) sampling,  the combination of big data and complex models conspire to produce significant challenges for the traditional MCMC algorithms. We review modern algorithmic developments addressing the latter and compare their performance using  numerical experiments.

\end{abstract}


\maketitle

%
%
\tableofcontents

\section{Introduction}

The data science revolution has led to multiple pressure points in statistics. A statistical sample from a large population exhibits, in the 21st century,  very different characteristics than what one would have seen merely a few years ago. The  ubiquitous and almost continuous recording  of many of our activities  has made it relatively easy to collect enormous amounts of information that require analysis and interpretation. This  drastic increase in data volume  imposes  sober re-evaluations of most classical  approaches to statistical inference. 

The computational side of a  Bayesian statistician's  toolbox is perhaps most challenged by these developments. The impetus of Bayesian statistics that has been felt since the early 1990s has been given by the spectacular advances in computation, especially those around Markov chain Monte Carlo (MCMC) sampling. 
Thanks to  methods in this class of algorithms, the statisticians have been liberated to think freely about the Bayesian model components used for a given problem, without worrying about the mathematical intractability of the analysis. 

Indeed, given a data set $\by$, most of the pairings of a sampling density, $f(\by\vert\bth)$, and a prior, $p(\bth)$, result in a posterior distribution 
\beq
\pi(\bth\vert\by) =  {p(\bth)f(\by\vert\bth) \over \int p(\bth)f(\by\vert\bth)d\bth}
\label{post}
\eeq
that  cannot be analyzed directly, usually because the denominator in \eqref{post} cannot be computed analytically. The latter fact impedes the calculation of quantities of interest related to $\pi$, most of which can be expressed as 
\beq
I =\int h(\bth)\pi(\bth\vert\by) d\bth,
\label{mcprob}
\eeq
for some function $h$ that is determined by the question of interest. 
For instance, if $\bth$ is univariate and we let  $h(\bth)=\bth^r$ in \eqref{mcprob}, then  $I$ is equal the $r$-th moment of $\pi$, or $h(\bth)=\one_{(-\infty,t]}(\bth)$ leads to the cumulative distribution function (cdf) of $\pi$ at a point $t$.

The classical Monte Carlo method, devised by \cite{von1951monte} at the middle of the twentieth century,  relies on sampling independently $\{\bth_1,\ldots, \bth_m\}$ from distribution $\pi$ and approximating $I$ with 
\beq
\hat I={1 \over m} \sum_{k=1}^m h(\bth_k).
\label{mcsol}\eeq
However, the unknown constant in \eqref{post} creates a knowledge gap that an MCMC algorithm closes by constructing and running a Harris-recurrent, $\pi$-irreducible,  aperiodic  Markov chain whose stationary  distribution is exactly the posterior distribution $\pi(\bth\vert\by)$. The values taken by the chain  make up the samples $\bth_1,\ldots, \bth_m$. A couple of issues emerge immediately. First, because $\pi$ is the chain's stationary distribution, the samples will be approximatively distributed with $\pi$ only after the chain has entered its stationary regime.   Second, due to the Markov property, the samples are typically positively correlated \citep[although see][for exceptions]{FGR, craiu2005multiprocess}
which reduces the amount of information they contain about $\pi$. To see that, let us imagine the extreme case in which the $m$ samples are perfectly correlated,  in which case they would provide very little information about $\pi$.

The success  MCMC sampling had in boosting the use of Bayesian models is largely due to the ease of implementation of some of its most popular algorithms. For instance,  the Metropolis-Hastings algorithm  (henceforth, MH)
\citep{metropolis1953equation, hastings1970monte} can be implemented using the following recursive procedure. 

\begin{description}
\item [Step 0] Initialize the Markov chain at $\bth_0$ and choose a {\it proposal} density $q(\cdot \vert \zeta)$ which may or may not depend on $\zeta$. 

\item[Step t] At the $t$-th step ($1\le t \le m-1$) do:
\begin{description}
\item[PR] Draw proposal $\omega_t$ from the  density $q(\cdot \vert\bth_{t})$;
\item[AR] Set
\[
\theta_{t+1}=\left \{
\begin{array}{cc}
\omega_t & \mbox{ with probability } \alpha_t \\
\bth_{t} & \mbox{ with probability } 1-\alpha_t \\
\end{array}
\right.
\]
where 
\beq
\alpha_t = \min \left \{ 1, {\pi(\omega_t\vert\by)q(\bth_{t}\vert\omega_t) \over \pi(\theta_{t}\vert\by)q(\omega_t\vert\bth_{t})} \right \}.\label{accprob}
\eeq
\end{description}
\end{description}

Because of the form of the acceptance probability \eqref{accprob}, its calculation is not prevented by the unknown denominator in \eqref{post}. Nevertheless, computation of \eqref{accprob} hinges on the ability to calculate the sampling density $f(\by\vert\bth)$ for any parameter value $\bth$ and to be able to do it $m$ times.  The  challenges posed to Bayesian computation by the modern data and modelling environment have their roots in this implicit assumption.

In very broad strokes, one can speak of two main challenges in modern Bayesian computation. The first one concerns the computational price of calculating a likelihood when the data is massive, say of order $N$ (think of $N$ as being of the order of hundred of millions or even billions). Even in the tame case of iid observations, in order to know \eqref{accprob} we will have to compute a likelihood (or sampling density) that involves $N$ terms and this will have to be repeated each time a new MCMC sample is produced. The cumulative cost is unsustainable as a single MCMC iteration can take days. A second challenge emerges when the model's complexity keeps up with the data volume and yields an intractable likelihood so that \eqref{accprob} simply cannot be computed analytically. Finally, a meta-challenge  appears in Bayesian analyses that merge massive data with intractable models. 

This paper discusses MCMC-adjacent methodology that is used to alleviate the pressure posed on Bayesian computation by the above challenges. Space constraints impedes the presentation of  details and variants, but in all cases we present the main ideas and  refer the  interested reader to the relevant literature. In order to gauge their computational efficiency, we run numerical experiments where, using publicly available software packages, the algorithms are implemented on two statistical models.  

In the next section we describe in more detail the challenges we just described and Section 3 summarizes some of the   solutions proposed to address them.   Numerical experiments meant to illustrate and compare different algorithms are reported in Section 4. The paper ends with comments and discussion of future directions for research.

\section{Modern challenges for Bayesian computation: Massive Data}

Consider data $\by$ collected on $N$ independent items so that $\by=\{y_1,\ld,y_N\} \in \X^N$ and denote by $f(\by\vert\bth)$  the sampling distribution which depends on  parameter $\bth \in \Theta \subset \RR^d$. At each iteration of the MH sampler, one needs to compute $f(\by\vert\omega_t) = \prod_{k=1}^N f(y_k\vert\omega_t)$ where $\omega_t$ is the proposal in \eqref{accprob}. Modern applications often rely on data that are large enough so that the repeated  calculation   of $f(\by\vert\omega_t)$ is impractical, even impossible. It is also not unusual for data size to be so large as to prohibit storage on a single machine, so that computation of the likelihood involves also repeated communication between multiple machines, thus adding significantly to the computational burden. 

Prompted by the obstacle of large data, computational Bayesians have designed a number of approaches to alleviate the problem. Two general ideas are currently standing out in terms of popularity and usage: divide-and-conquer (DAC) strategies and subsampling with minimum loss of information. 

\subsection{Divide and Conquer}

The DAC approach is based on partitioning the sample into a number of sub-samples, called batches that are analyzed separately on a number of workers (CPUs, GPUs, servers, etc). After the batch-specific estimates about the parameter of interest are obtained, the results are combined so that the analyst recovers a large part of, ideally all, the information that would have been available if the whole sample were analyzed in the usual way, on a single machine. While this idea seems applicable in a wide range of scenarios, there are a couple of constraints that restrict its generality. First, the procedure is computationally effective if it is designed to minimize, preferably eliminate, communication between the workers before combining the batch-specific results. Second, it is often difficult to produce an accurate assessment of the resulting loss of information at the combining stage. 
Some of the first proponents of  DAC for MCMC sampling are  \cite{emb}, \cite{cons}, and
\cite{weierstrass}. In their approach, the {\it subposterior} distribution corresponding to the  $j$th batch, is defined as 
\beq
\pi^{(j)} (\bth\vert\by^{(j)}) \propto f(\by^{(j)}\vert\bth)[p(\bth)]^{1/J}
\label{eq:cmc}
\eeq
where  $f,p$ are as in \eqref{post}, $\by^{(j)}$ is the data that was assigned to batch $j$, $1 \le j \le J$,  and  $J$ is the total number of batches. With this choice, one immediately gets that $\prod_{j=1}^J \pi^{(j)} \propto \pi(\bth\vert\by)$. 
Both \cite{emb} and \cite{cons} consider ways to combine  samples from the subposteriors $\pi^{(j)}(\bth)$, $1\le j \le J$, in situations in which all posteriors,  batch-specific and full data ones, are Gaussian or can be approximated by mixtures of Gaussians. in this case, one can demonstrate that a weighted average of samples from all the $\pi^{(j)}$'s have density $\pi$. The use of the Weierstrass transform for each posterior density, proposed in \cite{weierstrass}, extends the range of theoretical validity beyond Gaussian distributions. The authors also establish error bounds between the approximation and the true posterior.  \cite{nemeth2018merging} use a Gaussian process (GP) approximation of each subposterior. Once again, the Gaussian nature of the approximation makes recombination possible and relatively straightforward. Limitations of the method are strongly linked with those of GP-based estimation. For instance, when the sub-posterior samplers are sluggish, large MCMC samples might be needed which, in turn, make the calculation of the GP-based approximation very expensive. The idea of using the values of the sub-posterior at each MCMC sample is adopted also by  \cite{changye2019parallelising} who propose to define the subposteriors using $\pi^{(j)} \propto \{[p(\bth)]^{1/J}f(\by^{(j)}\vert\bth)\}^{\lambda_j}$. The scale factor $\lambda_j$ is used to control the uncertainty in the subposterior. Alternative ways to define the sub-posteriors are produced by \cite{entezari2018likelihood} who use $\pi^{(j)} \propto p(\bth)[f(\by^{(j)}\vert\bth)]^J$. The intuitive idea is to "match" the size of the original sample and  the batch-specific one. Their approach has been applied successfully to BART \citep{chipman2010bart,pratola2016efficient} models. 

\subsection{Subsampling}

Subsampling approaches are mostly developed under two assumptions. The first one is that with massive data one expects a certain amount of redundancy, so it is possible to the same likelihood when we eliminate a  proportion of the sample as long as the remaining observations are properly weighted. A simple illustration is one in which $R$ observations are identical, so that $R-1$ of them can be taken out of the likelihood calculation if the term corresponding to the remaining one is raised to power $R$.  The second idea is that one might use only a small percentage of the sample to find accurate (e.g., unbiased) approximations of the quantities needed to run an MCMC sampler.  For instance, in the case of a MH sampler, the pseudo-marginal approach of \cite{andrieu2009pseudo} demonstrates that the stationary distribution of the chain is the same when   the likelihoods  involved in \eqref{accprob} are replaced  with unbiased estimators. The pseudo-marginal idea has largely impacted  the methods based on subsampling for MCMC. While some divide the latter into exact and approximate, we will refrain from using similar taxonomy because, in our opinion, all subsampling MCMC methods introduce some level of approximation into the computation of interest.  

Early efforts include those of \cite{korattikara2014austerity} and 
\cite{bardenet2014towards}  who propose estimating the acceptance probability \eqref{accprob} using only a random subset of the data. The latter authors demonstrate that, with probability higher than a threshold set in place by the user, their method yields estimates that are equal to the one produced by the full data likelihood. However, one does not know in advance the size of the sample needed at each iteration and thus must be able, in principle, to access most of it  at all times. A review of early subsampling methods can be found in \cite{bardenet2017markov}. 

\subsubsection{Coresets}
The process of establishing which sample points are redundant must have theoretical backing, lest it leads to a very different posterior distribution without any hope to control or assess the error incurred. The coreset approach of  \cite{campbell2019automated} offers theoretical guarantees about the quality of the approximation resulting from sample reduction. Consider the loglikelihood obtained from $N$ iid observations
\beq
 l(\bth\vert \by)= \sum_{i=1}^N l_i(\bth\vert y_i), 
\label{loglik}
\eeq
where $l_i(\bth\vert y_i)= \log f(y_i\vert\bth)$. The aim of the coreset method is to find a set of weights $\{w_i: \; 1\le i \le N\}$, most of them zero, so that \beq
\|\Lambda(\bth\vert \bfW,\by)- l(\bth\vert\by)\|\le \eps \| l(\bth\vert\by)\|,
\label{obj}
\eeq
 for all $\bth \in \Theta$, where $\bfW =(W_1,\ldots, W_N)$ is the vector of weights, and $\Lambda(\bth \vert \bfW, \by)=\sum_{i=1}^N W_il_i(\bth\vert y_i)$ is the weighted log-likelihood of the coreset. 
The weights found by \cite{campbell2019automated} are defined as 
\beq
W_i = {\sigma \over \sigma_i}{M_i \over M},
\label{csw}
\eeq
where \beq\sigma_i=\sup_{\bth \in \Theta} \left \| {l_i(\bth\vert y_i) \over  l(\bth\vert \by)} \right\| \label{sens}\eeq is called the {\it sensitivity} of the $i$-th observation,  $\sigma =\sum_{i=1}^N \sigma_i$, $M$ is the size of the coreset and $(M_1,\ldots,M_N) \sim Multi\left( M, \left \{{\sigma_i \over \sigma}: \; 1\le i \le N \right \} \right)$ are multinomial draws. One can think of the sensitivity in \eqref{sens} as a measure of the influence of the $i$-th observation on the whole likelihood as $\theta$ varies. As expected, the algorithm will retain observations that correspond to relatively higher likelihood values, but more importantly, it allows some evaluation of the error incurred when the sample is reduced. The ideas that led to the weights in  \eqref{csw} illustrate the general principles of the approach, but improvements are possible when one considers other norms in \eqref{obj} and \eqref{sens}. For instance,  \cite{campbell2019automated} consider the $l_i$'s as vectors in a Hilbert space, link the norm to the inner product in the space and include directionality in the selection of the coresets. The latter allows replacing the  simultaneous selection of the coreset elements by a more intuitive procedure in which samples are sequentially added to minimize the residual error.    In Section 3 we implement the coreset approach for logistic regression as presented in \cite{huggins2016coresets}. For this model, the coreset is build along the principles delineated above and requires some specific tuning. The parameter space is taken to be an Euclidian ball of radius $R$ which is a reasonable working assumption in the case of a logistic regression with standardized covariates. The sensitivity measure for each point is modified after $K$-clustering the entire sample.  A measure of spread within each cluster is used to construct upper bounds for  the sensitivity of each point in the sample. The intuition guiding this choice is that clusters whose data vectors are tightly bundled together will be  well represented in the coreset by only a few points, while clusters with more spread will need to contribute more points.   Overall, the coreset construction is intuitive and offers many possible directions for future research. The biggest challenge is the evaluation of approximating error induced in the posterior when replacing the full sample by the coreset, although some promising initial results exist \citep{manousakas2020bayesian}.

\subsubsection{Random Subsampling}
One can think of  coreset subsampling as a static approach, in the sense that the subsample is selected once and the Bayesian analysis is subsequently conducted using the coreset in lieu of the original sample.  A more dynamic approach is considered by \cite{quiroz2018speeding} who propose to use all the data for inference, just not at once. Their  idea is to use a different subset of the data each time the MCMC chain is updated. For instance, in the case of an MH sampler, a different subset of individuals will contribute to the likelihood needed in the calculation of \eqref{accprob}, at each iteration.  Following the development of pseudo-marginal strategies, \cite{andrieu2015convergence} studied  the convergence properties of an MH or a random walk Metropolis sampler in which the likelihood in \eqref{accprob} is replaced by an unbiased estimator. They have shown that 
the efficiency of the MCMC sample increases when the variance of the unbiased estimator decreases.

The use of subsampling within MCMC  proposed by  \cite{quiroz2018speeding} can be applied quite generally and it is attractive because it addresses both the construction of the unbiased estimator for the likelihood and the reduction of its variance.   

Given a random subsample of $\by$ of size $m$, 
$\by_{\bu}=\{y_{u_1},\ldots, y_{u_m}\}$, where 
$\bu= \{u_1,\ldots, u_m\}$ are iid random variables uniformly distributed over 
$\{1,\ldots,N\}$, the following estimator
\beq
l_{m}(\bth|\by_{\bu}) = {1\over m}\sum_{k=1}^m l_{u_k}(\bth|y_{u_k}),
\label{est1}
\eeq
is unbiased for the average log-likelihood ${1\over N}l(\bth|\by)$. 
However, it usually has a large variance, subjecting the pseudo-marginal chain that uses \eqref{est1} instead of the full-sample likelihood in \eqref{accprob} to an increased risk of poor mixing, since an unusually high value of the likelihood at the current state of the chain will make it unlikely to  accept a proposal.  A reduction in variance is desirable and can be achieved via control variates (CV),
$\bfq (\bth) = \{q_{1}(\bth),\ldots, q_{N}(\bth)\}$, and via a modified estimator of \eqref{est1},
\beq
\tilde l_{m}(\bth|\by_{\bu}, \bfq) = \sum_{i=1}^N q_i(\bth) + {N\over m}\sum_{k=1}^m (l_{u_k}(\bth|y_{u_k}) - q_{u_k}(\bth)). 
\label{est2}
\eeq

The notation implies that $\bfq$ might change with $\bth$. Indeed, when the likelihood is unimodal,
the construction of the control variate follows \cite{bardenet2017markov} who use for each $\bth$, a Taylor series expansion of $l(\theta|\by)$ around a fixed point, $\bth^*$ which is a point centrally located in $\Theta$ ( e.g. the maximum likelihood estimate) so that for all $1\le i \le N$
\beq
q_i(\bth)= l_i(\bth^*|y_i)+ (\bth -\bth^*)^T {d \over d\bth} l_i(\theta^*|y_i) +
{1 \over 2} (\bth -\bth^*)^T  {d^2 \over d\bth^2} l_i(\theta^*|y_i) (\bth -\bth^*).
\label{cv1}
\eeq
This control variate is called {\it parameter expanded} by \cite{quiroz2018speeding} because it is obtained using an expansion in the parameter space. With this modification, running a MH chain for, say, $M$ iterations requires the evaluation of $N+mM$ item-specific likelihood terms, $l_i(\bth|y_i)$, for \eqref{est2}   and $MN$ for \eqref{loglik}. This can translate into significant reduction of computation effort when $m<<N$. 

When the likelihood is multimodal or the Taylor approximation is poor when $\bth,\bth^*$ are  distanced, the authors discuss an alternative construction that identifies  a number of  centroids  
$\by^*_1,\ldots, \by^*_r$ via clustering of the data and uses Taylor series expansions around each  centroid to define the so-called {\it data expanded} control variates. In Section 5 we implement the subsampling method with parameter and data expanded control variates.

The reduction in variance requires  a careful derivation in which the source of variability is provided by the finite distribution of the random vector $(u_1,\ldots u_m)$. The latter can be sampled at random at each iteration or one can use the ideas in \cite{deligiannidis2018correlated} and 
allow dependence  between the $u$'s in consecutive iterations to further reduce the variance of \eqref{est2}. In Section 5x we implement the subsampling method with random or correlated selection of indices, and parameter or data expanded control variates.

 Finally, we should also point out that while the estimators discussed are unbiased for the loglikelihood, this does not translate into an unbiased estimator for the likelihood itself. Therefore, an approximate correction term is applied to reduce the bias but does not dissolve it, which means that the pseudo-marginal theory cannot be applied mutatis mutandis in this case. Therefore, the target distribution of the chain is perturbed and one must assess the size of the error incurred. The authors produce a bound of the perturbation error and provide empirical evidence that their bound is conservative. Additional details and derivations can be found in  \cite{quiroz2018speeding}.


\section{Modern Challenges for Bayesian Computation: Intractable Likelihoods}

So far we have looked at the pressure posed by the size of the sample on Bayesian computation. However, there are other hurdles that accompany  a massive sample.  Often, large data imply more information which, in order to be used fully, requires a more complex  model. As data become richer and modellers are more ambitious, the likelihoods tend to get intractable, such as the ones used  in population genetics \citep{pritchard1999population,beaumont2002approximate}, groundwater studies \citep{cui2018emulator}, hurricane surges \citep{plumlee2021high}, or in climate change scenarios \citep{oyebamiji2015emulating}.  

At first sight, it can be surprising that Bayesian inference can still be conducted when the likelihood is intractable. The likelihood provides a crucial analytical link between any parameter value and the probability of observing a given data set. When such a link is not analytically tractable it will have to be inferred from simulations. Central to the latter approach is the ability to sample, given any value of the parameter, pseudo-data from the model. To provide an intuition, imagine that infinite computational resources are available. Then one can see that for any $\bth \in \Theta$ it is possible to simulate  enough pseudo-data sets to approximate at any degree of precision the distribution of the observed-data $f(\by_0|\bth)$, essentially filling the void left by the intractability of the likelihood. However, computational resources are not infinite so  ingenious ways are needed to reduce computational burden. We discuss here two algorithms,  Approximate Bayesian computation (ABC) and Bayesian Synthetic Likelihood (BSL), that have gained popularity in the statistical and, more generally, the scientific communities.

\subsection{Approximate Bayesian Computation (ABC)}
Our discussion of ABC will be brief, given the recent and excellent reviews 
of \cite{robert2014bayesian},
\cite{sisson2018overview}, and the  comprehensive handbook of ABC \citep{sisson2018handbook}.

The  ABC agorithm was initially 	proposed as an
 accept/reject sampler \citep{tavare1997inferring}.  Given any  $\bth^*$ sampled from the prior $p(\bth)$, it assumed that is possible to generate pseudo-data $\by$ from $f(\by|\bth^*)$. If the pseudo-data and the original data are close enough,  then the parameter $\bth^*$ is an approximate draw from the posterior $\pi(\bth|\by_0)$. Let us frame  next the ``close enough" and ``approximate draw" in precise mathematical terms and provide some justification for our choices.
 
Given $\eps >0$, a distance $d: \RR^p \times \RR^p \rightarrow \RR_+$ and summary statistic $S(\by) \in \RR^p$, the ABC algorithm has the following steps :
\begin{itemize}
\item[S1] Sample $\bth^* \sim p(\bth)$ and synthetic data $\by \sim f(\by|\bth^*)$
\item[S2] If $d(S(\by) , S(\by_0))\le \eps$ then accept $\bth^*$ as a sample from the approximate posterior $\pi_\eps(\theta|S(\by_0))$, the marginal (in $\bth$) of the joint distribution  
\begin{equation}
\pi_{\eps}(\theta,\by|S(\by_0))\propto p(\theta)f(\by|\theta)\one_{\{d(S(\by) , S(\by_0)) <\eps\}}.
\label{abc:t2}
\end{equation} 
\end{itemize}

If it is possible to have $\by = \by_0$ (for instance, if $\by_0$ is a discrete random variable with finite support)  we can choose $S(\by)=\by$ and $\eps=0$, then the approximate posterior is the true posterior, i.e.\ $\pi_\eps (\bth|\by_0) = \pi(\bth|\by_0)$. This is easier to see when both $\bth$ and $\by$ take discrete values. Then, one can easily see that 
\beqn
\pr(\bth=\bth_0)&\propto& p(\bth_0) \Pr(\by=\by_0|\bth=\bth_0)\propto \pi(\bth_0|\by_0)
\label{j1}
\eeqn 
where \eqref{j1}   holds because of the algorithm's construction with 
$S(\by)=\by$ and $\eps=0$. The above can be easily extended to the case when $S$ is a sufficient statistics.
In general, models with the level of complexity that requires ABC, will not have a low-dimensional sufficient statistics so the choice of $S$ is central to the performance of the ABC algorithm \citep{fearnhead2012constructing,marin2014relevant}. The accept-reject form of the ABC sampler makes it inefficient when the prior and posterior place most  of their mass on different regions of $\Theta$. 
Recognizing this, \cite{marjoram2003markov} proposed an  ABC-MCMC algorithm 
 which relies on building a Metropolis-Hastings  (MH) transition kernel, with state space $\{(\theta,\by) \in  \RR^q \times \calX^n\}$, proposal distribution at iteration $t$,  $ q(\theta|\theta_t)\times f(\by|\theta)$, and  target  
\begin{equation}
\pi_{\eps}(\theta,\by|S(\by_0))\propto p(\theta)f(\by|\theta)\one_{\{d(S(\by),S(\by_0))<\eps\}}
\label{abc:t2}
\end{equation} 
for which \eqref{accprob} can be computed exactly, because the intractable terms involving the likelihood, $f(\by|\bth)$, cancel out.  Alternatives to this include  
the pseudo-marginal approach of \cite{lee2012discussion}, or the sequential Monte Carlo implementation of \cite{sisson2007sequential, lee2012choice}, and \cite{filippi2013optimality}.

\subsection{Bayesian Synthetic likelihood}

Indirect inference was developed in  econometrics \citep{smith1993estimating,gourieroux1993indirect} for complex data models which are intractable, but can be sampled from. The central tenet is  that a complex model of interest, $f(\by|\bth)$,  can be well approximated   using a tractable sampling model $g(\by|\phi)$ where 
$\dim(\phi)> \dim(\bth)$. In other words, the complex model can be approximated by a simpler model whose parameter space is of larger dimension and has a tractable likelihood.  For instance, Bayesian  estimation of $\bth$ is possible if one estimates its functional connection with $\phi$ \citep[see, for instance,][]{gallant2009determination}.

The BSL algorithm \citep{price2018bayesian} relies on the synthetic likelihood (SL) approximation of  \cite{wood2010statistical} which falls squarely in the class of indirect inference methods. The idea hinges on the assumption that the conditional distribution  $p(S(\by)|\bth)$  is well approximated by a multivariate Gaussian $\CN(\mu(\bth),\Sigma(\bth))$ whenever $\by \sim f(\by\vert \bth)$.  The SL is defined as 
$\sil(\bth)=n(S(\by); \mu(\bth), \Sigma(\bth))$ where $n(\cdot; \mu, \Sigma)$ is the density of a multivariate normal with mean $\mu$ and variance $\Sigma$. 
One can estimate  $\mu(\bth), \Sigma(\bth)$ numerically, for any $\bth$. Given $\bth$, it is enough to  repeatedly sample pseudo-data $\by_j\sim f(\by|\bth)$ and compute $S(\by_j)$ for  
$1\le j \le K$ and then estimate $\hat\mu(\bth)= {1\over K} \sum_{j=1}^K S(\by_j)$ and $\hat \Sigma(\bth)= \mbox{SamVar}(\{S(\by_j):\; 1\le j \le K\})$ where SamVar is the sample variance of the computed statistics. BSL is then based on the approximation
$\pi_{BSL}(\bth|S(\by_0)) \propto p(\bth)\sil(\bth|S(\by_0))$ which can be explored via MCMC sampling using the following update rule at iteration $t>0$:

\begin{description}
\item[PR] Generate $\bth^* \sim q(\cdot|\bth_{t})$,
 estimate $\hat \mu_{\bth^*},\hat \Sigma_{\bth^*}$ from $K$ pseudo-data  $\{\by_{j} \sim f(\by|\bth^*): \; 1\le j \le K\}$, and compute  $SL(\bth^*) = \CN(S(\by_0);\hat \mu_{\bth^*},\hat \Sigma_{\bth^*})$.

\item[AR] Set $\bth_{t+1}=\bth^*$ with probability 
$\alpha = \min\left(1,\frac{p(\bth^*)SL(\bth^*) q(\bth_{t}|\bth^*)}{p(\bth_{t})SL(\bth_{t})q(\bth^*|\bth_{t})}\right)$ and $\bth_{t+1}=\bth_{t}$ otherwise.
\end{description}

\section{Double Jeopardy}

The separate treatment of the challenges brought by the big data or  intractable models is artificial and we anticipate that, more and more, the two challenges will have to be met simultaneously.  Since the use of MCMC within  ABC or BSL procedures requires repeated generation of pseudo-data of the same size and complexity as the observed ones, it incurs unmanageable computational costs when the data are massive or the data generating procedure is expensive. 

Some of the methods described within the first challenge are amenable to being used in combination with ABC or BSL. For instance, DAC strategies can be used for an intractable model if  each worker runs a separate  ABC MCMC sampler for each  batch of data.  The obvious caveat is the difficulty of ascertaining the loss of information after the merging stage. Unfortunately, more generalizable methods like those used for  subsampling cannot be used within ABC or BSL. 

A strategy  customized to ABC and BSL samplers with large or complex data is proposed by \cite{levi2021finding}. We describe here a  variation of their approach which combines finite adaptation  ideas and pre-sampling of the proposals. Assuming that a MH transition kernel  is used to  implement  ABC MCMC or BSL MCMC, the first $B$ samples are used to tune the proposal distribution. For instance, if a Gaussian proposal is used then its covariance matrix can be estimated using methods proposed by \cite{Haario:2001kx}, \cite{adaptex} or, in the case of multimodal targets, by \cite{craiu-jeff-yang} or \cite{pompe2020framework}.
The  computational effort is reduced because we rely on a set of  proposals that are generated in advance. This allows an embarassingly parallel procedure that benefits from the use of  multiple workers. The preprocessed draws are collected in reference set 
$\calZ =\{(\xi_h,s_h=(s_h^{(1)},\ldots, s_h^{(m)})^T): 1\le h \le H\}$ where for each parameter value $\xi_h$  generated from the proposal distribution, we sample $m$ pseudo-data $\bw_h^{(1)},\ldots,\bw_h^{(m)} \stackrel{iid}{\sim} f(\bw|\xi_h)$ and  set $s_h^{(j)}=S(\bw_h^{(j)})$ for all $1\le j \le m$. Note that the set $\calZ$ is generated independently of the chain. 

We illustrate here the use of $\calZ$ to run the  ABSL sampler.  If the chain's proposal at $t$th iteration,  $\bth^*$, is identical to one element, say $\xi_h \in \calZ$, and $m$ is large, then we would not need to generate 
$\by_1,\ldots,\by_m \sim f(\by|\bth^*)$ since we already have the corresponding pseudo-data statistics vectors $s_{h}$ which can be used to estimate $\mu(\bth^*)$, $\Sigma(\bth^*)$ and thus $\sil(\bth^*)$.  While the intuition is attractive,  it is impractical to faithfully implement it. For instance, using a large value for  $m$ when creating $\calZ$ might still be too costly and an exact match with an element in the reference set is unattainable when the parameter space is continuous. However, if
$\calZ$  contains enough $\xi$-values  that are close enough to $\bth^*$, one can still use them for  estimating $\sil(\bth^*)$.  \cite{levi2021finding} build the reference set with $m=1$ and propose the use  of $K$-nearest neighbours (kNN) estimators for  
$\mu(\bth^*)$, $\Sigma(\bth^*)$
\begin{equation}
\begin{split}
\tilde \mu(\bth^*) & = \frac{\sum_{h=1}^H [ W_{h}(\bth^*) {1\over m} \sum_{j=1}^m s_h^{(j)}]}{\sum_{h=1}^H W_{h}(\bth^*)},\\
\tilde \Sigma(\bth^*) & =  \frac{ \sum_{h=1}^H [W_{h}(\bth^*) {1\over m}\sum_{j=1}^m (s_h^{(j)}-\hat\mu_{\bth^*})(s_h^{(j)}-\hat\mu_{\bth^*})^T ]}{\sum_{h=1}^H W_{h}(\bth^*)}.
\end{split}
\label{w-est}
\end{equation}  
where $W_h (\bth^*)= 1$ or $W_{h}(\bth^*)=1-\|\xi_h -\bth^*\|/\|\xi^*-\bth^*\|$ and 
$\xi^*=\max_{\xi \in \calZ}\|\xi - \bth^*\|$, i.e. is the point in $\calZ$ that is furthest away from $\bth^*$.
If $H$ is large, it is likely that most of its elements will contribute little or not at all to the estimators \eqref{w-est}. Instead of summing over all the $H$ elements in $\calZ$, it is then advisable to use only the  $K$  $\xi$'s that are closest to $\bth^*$, where $K$ is user-defined and depends on the available computational power. In our numerical experiments we have used $W_h=1$ for all 
$1\le h \le K$ and $K=\lfloor\sqrt{H}\rfloor$. Clearly, the estimators in \eqref{w-est} are consistent due to the properties of kNN estimators, but are not unbiased, so  pseudo-marginal arguments cannot be invoked to justify the approach. Validity is demonstrated theoretically by showing that the perturbation induced when using the modified transition kernel  can be controlled using the user-specified tuning parameters  of the sampler \citep[see section 6 in][for details]{levi2021finding}. 

A similar approach is used by \cite{levi2021finding} for the ABC MCMC chain that targets  the marginal posterior density of $\bth$ resulting from \eqref{abc:t2}, $\pi(\bth|S(\by_0))\propto p(\bth)\Pr(d(S(\by) ,S (\by_0))<\eps)|\bth)$. Instead of using an unbiased estimator for $\Pr(d(S(\by) ,S (\by_0))<\eps)|\bth)$ which would require multiple pseudo-data generated from $f(\by|\bth)$, they construct the kNN-based estimator from the collection $\calZ$.

In the next section, we compare numerically the methods discussed so far using a couple of examples.

\section{Numerical experiments}
In this section we present the performance of the discussed algorithms on two models: logistic regression and stochastic volatility.   We compare the accuracy and computational efficiency of the described methods with a couple of benchmark MCMC algorithms that are widely known to perform very well in these cases. 
\color{black} Specifically, we measure the perfomance of the methods presented in this paper against  the Polya-Gamma (PG) sampler  \citep{polson2013bayesian} for the logistic regression, and the  sequential ABC (ABC SMC) of \cite{sisson2007sequential, lee2012choice}
for the stochastic volatility  model.   
The former is customized for logistic regression and for the latter the length of $\eps$ sequence is set at $15$.   

\color{black}

\subsection{Description of the simulation settings}
The following variations of the algorithms described in previous sections are implemented. 

\begin{description}
    \item[PG\_DAC\_J] DAC algorithm with PG sampler that follows the setup in  \cite{cons} using \eqref{eq:cmc}. The samples from each batch are combined proportionally to the inverse covariance matrices. $J$ denotes the number of batches.
    \item [RW\_SS] Subsampling using \cite{quiroz2018speeding} with a Random Walk (RW) transition kernel. There are four variations corresponding to pairing parameter or data expansion with random or correlated  index selection.
     \begin{description}
       \item [RW\_SS\_P\_R\_m] - Parameter expansion and random index selection
       \item [RW\_SS\_D\_R\_K\_m] - Data expansion and random index selection
       \item [RW\_SS\_P\_C\_m] - Parameter expansion and correlated index selection, the correlation $\rho$ is set at $\rho=0.9999$
       \item [RW\_SS\_D\_C\_K\_m] - Data expansion and correlated index selection, the correlation $\rho$ is set at $\rho=0.9999$
     \end{description}
     Note that $m$ and $K$ indicate the number of observations that will be evaluated with the actual log-likelihood and the number of clusters respectively.
     \item[RW\_CO\_K\_f] Coreset method for logistic regression proposed by \cite{huggins2016coresets}. The number of clusters and proportion of non-zero weights out of $N$ are specified by $K$ and $f$, respectively. Note that the radius $R$ is calculated from the average sum of squared distances within each cluster as suggested in \cite{huggins2016coresets}. The coreset is used with a random walk Metropolis (RWM) sampling algorithm. 
            \item[RW\_ABC] ABC MCMC algorithm using a RWM transition kernel for target \eqref{abc:t2}. Only one pseudo data set is generated for each proposal $\bth^*$.
     \item [RW\_AABC] Approximate ABC MCMC algorithm proposed by \cite{levi2021finding}.  Proposals from the history of the chain are used to estimate the likelihood using the k-nearest-neighbour approach with uniform weights. Only one pseudo data set is generated at every iteration. 
     \item [RW\_BSL\_m] BSL MCMC  algorithm with a RWM transition kernel. The distribution of the summary statistics is approximated by a Gaussian. The mean and covariance of the distribution is estimated by generating $m$ pseudo data sets at each proposed $\bth^*$.
     \item [RW\_ABSL] Approximate BSL MCMC algorithm proposed by \cite{levi2021finding}. Past results are used to estimate the mean and covariance of the summary statistics distribution using k-nearest-neighbour approach with uniform weights. Only one pseudo data set is generated at every iteration. See more details in the Supplementary Materials.
\end{description}

With the exception of  PG and ABC SMC, all the approximate samplers rely on a random walk Metropolis (RWM) kernel with Gaussian proposals to ensure consistency and comparability.   The RWM kernels used here benefit from a finite-adaptation strategy, in which the covariance of the proposal is modified using the method of  \cite{Haario:2001kx}, during the first $B$ iterations that make up the  burn-in period,
   and are kept fixed after that. The ABC, AABC and ABC SMC samplers depend on the threshold $\eps$ and the  ingredients needed to compute the distance $d$ in \eqref{abc:t2}, and are determined following preliminary simulations, as detailed in \cite{levi2021finding}.  Additional details about each sampling design are provided in the Supplementary Materials.

For standard MCMC samplers, their performance comparison is often reported in terms of the effective sample size (ESS) per second of central processing unit (CPU) time, denoted ESS/cpu. The ESS is interpreted as the number of independent samples that would yield the same variance of the Monte Carlo estimator.  A higher ESS value indicates a more efficient MCMC sampling algorithm, since it has been directly linked with the algorithm's computational uncertainty 
\citep[e.g.,][]{gong2016practical,vats2019multivariate}. The CPU time directly measures the computational cost in seconds so ESS/cpu can be interpreted as a sampler's speed of generating information about the target. 

All the samplers discussed here will target a distribution different than the posterior of interest. Thus, in order to fully compare these sampler, one must consider the errors incurred because of this shift.  Therefore, in addition to metrics designed to measure the efficiency of a regular MCMC sampler, such as ESS/cpu, we also use  $R=50$ independent replicates to produce  estimates of Monte Carlo bias and variance.   This led us to two  measures of efficiency that are used to convey the performance of each method: the root mean square error (RMSE), and  the ESS/cpu. 

To fix the notions, let $\bth^{rs}_{(t)}$ represent the posterior samples from replicate $1\le r \le R$, iteration $B\le t \le M$ (only draws obtained after burn-in are retained) and parameter component $1\le s\le d$. Similarly, $\tilde \bth^{rs}_{(t)}$ are posterior draws from the benchmark chain (only draws obtained  after the burn-in period are retained). We also let $\bth_{true}^s$ denote the true parameter value that was used to generate the data. 
The following quantities are used for comparing computational efficiency
\begin{equation*}
\begin{split}
& \mbox{Bias}^2 = \mbox{Mean}_s\left(\left( \mbox{Mean}_{tr}(\bth^{rs}_{(t)}) - \bth_{true}^s \right)^2\right), \\
& \mbox{VAR} = \mbox{Mean}_s(\mbox{Var}_r(\mbox{Mean}_t(\bth^{rs}_{(t)}))), \\
& \mbox{RMSE} = \sqrt{\mbox{Bias}^2 + \mbox{VAR}},
\end{split}
\end{equation*}
where $\mbox{Mean}_t(a^{st})$ is defined as the average of $\{a^{st}\}$ over index $t$ and, similarly, $\mbox{Var}_t(a^{st})$ and $\mbox{Cov}_t(a^{st})$ denote the sample variance and covariance, respectively. 


Using the {\tt{coda}} library in R we compute ESS for each replicate and parameter's component $\mbox{ESS}^{rs}$. Letting $CPU^{r}$ denote the total CPU time used for producing the MCMC samples in replicate $r$, we define ESS/cpu as:
\begin{equation*}
\begin{split}
& \mbox{ESS/cpu} = Mean_{rs}(\mbox{ESS}^{rs}/CPU^r). 
\end{split}
\end{equation*} 
Note that we consider the average over all  parameters and replicates.  Generally, a sampler with a higher ESS/cpu  is preferred because it yields a higher amount of information per unit of time. The ABC SMC sampler  produces independent draws so its ESS is equal to the number of particles.

Finally, in order to frame the comparison in terms of unit-free measures, we report the  performance relative to the benchmark samplers. This means that once we compute the RMSE for say method A, RMSE$_A$, we report instead $\mbox{RMSE}_A/\mbox{RMSE}_{Bench}$, where the denominator is the benchmark sampler's RMSE. Similarly, we also report the relative ESS/cpu performance.

\subsection{Logistic Regression}
This set of simulations contains the standard setting for the logistic regression model. The design $N\times d$ matrix $X$ is generated by simulating each variable independently from $Unif(0,1)$. The left most column is a column of $1$s (intercept). For $i=1,\ldots,N$, $Y_i$ is Bernoulli with 
$\pr(Y_i=1)=\mbox{logistic}(x_i \cdot \bth_{true})$. We considered two values for the sample size $N$: $1,000$ and $10,000$; and two sets of parameters:
\begin{itemize}
\item $d=2$ with the true parameter of $\bth_{true} = (-2,2)$
\item $d=10$ with the true parameter of $\bth_{true} = (-2,2,-3,4,1,2,-3,-4,2,1)/3$
\end{itemize} 
We set the prior distribution to be $p(\bth)\sim \mathcal{N}(0,4I_{d})$,  where $\bth \in R^d$ and $I_d$ is $d\times d$ identity matrix. 

All the samplers  are run for $M=55,000$ iterations with burn-in set at $B=15,000$. 

\begin{figure}[!ht]
\begin{center}
\includegraphics[scale=0.40]{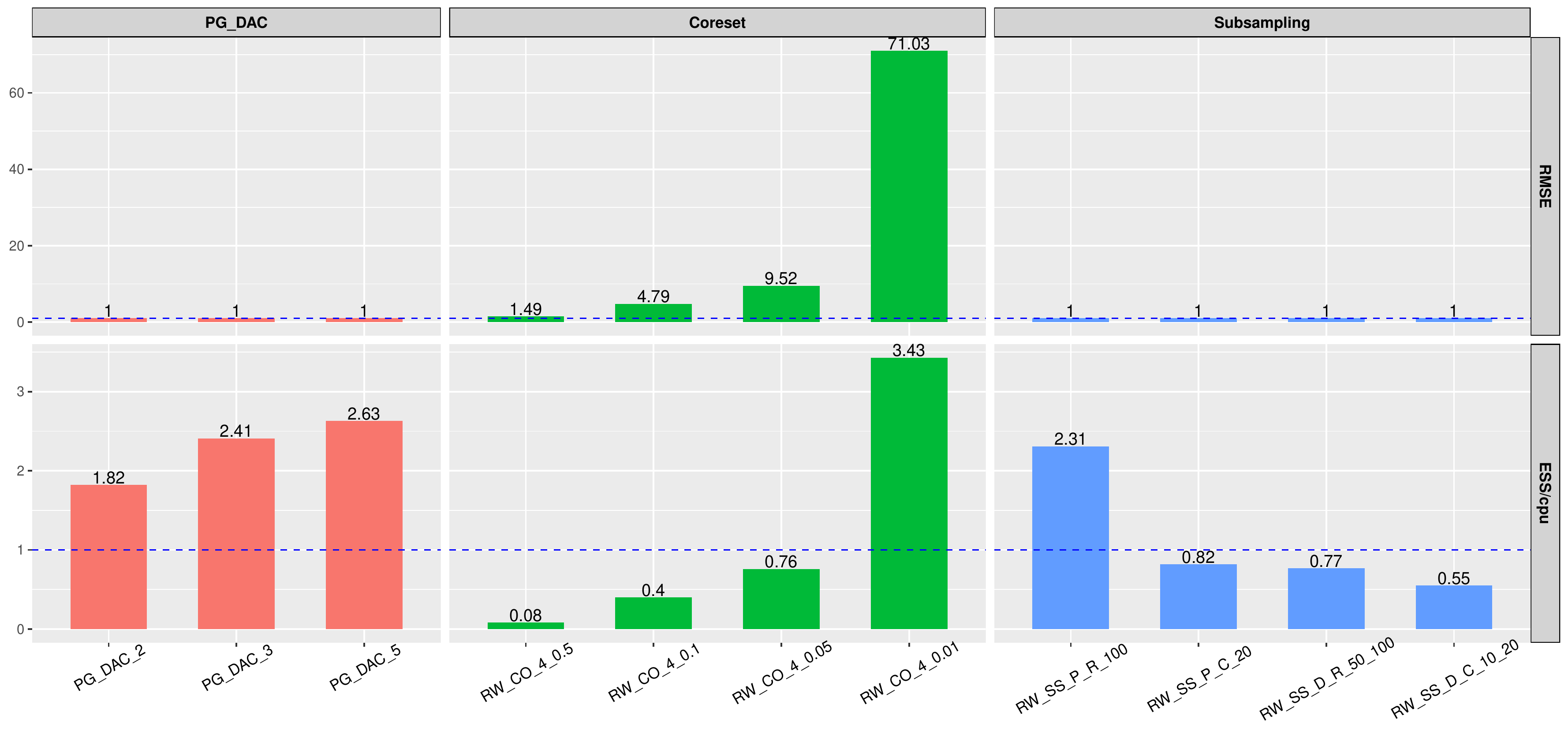}
\caption {Logistic model: Relative RMSE ({\em top row}) and ESS/cpu ({\em bottom row}) when $N=10,000$ and $d=2$ for   DAC-based samplers ({\em left column}),  Coreset-based samplers ({\em center column}), and Subsampling-based samplers ({\em right column}).}
\label{fig:log.sim.10k.2}
\end{center}
\end{figure}

\begin{figure}[!ht]
\begin{center}
\includegraphics[scale=0.40]{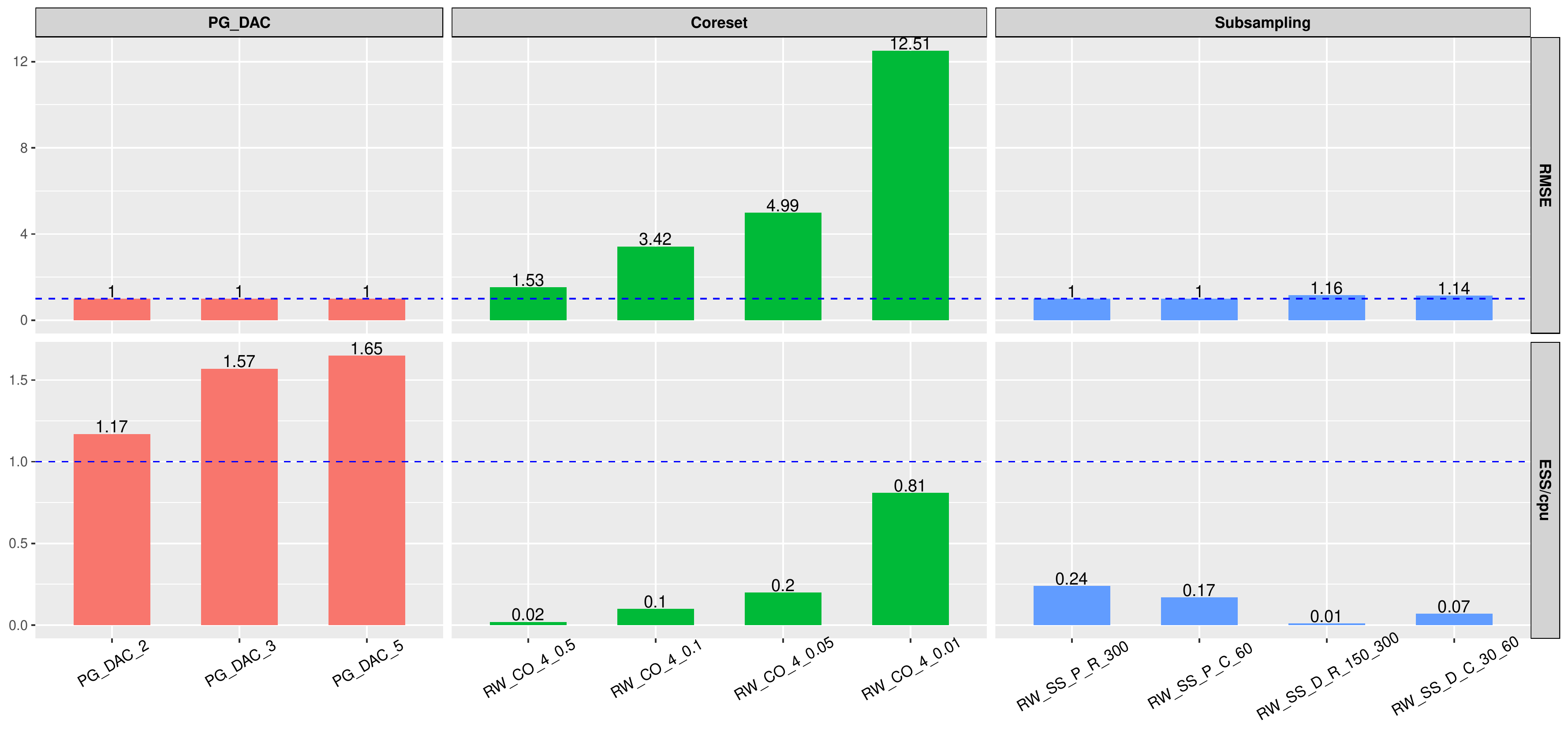}
\caption {Logistic model: Relative RMSE ({\em top row}) and ESS/cpu ({\em bottom row}) when $N=10,000$ and $d=10$ for   DAC-based samplers ({\em left column}),  Coreset-based samplers ({\em center column}), and Subsampling-based samplers ({\em right column}).}
\label{fig:log.sim.10k.10}
\end{center}
\end{figure}

For the DAC sampler we consider three values for the number of batches $J=2,3,5$. We set $K=4$ for the Coreset method and compare four values for the fraction $f=0.5,0.1,0.05,0.01$. 
\color{black}Finally, the values of the tuning parameters for the subsampling method were also variable. Specifically, the  the number of data clusters, $K \in \{10,50\}$, and the size of the subsample, $m\in \{20,100\}$. Generally, $K$ and $m$ will depend on the sample size $N$ and parameter dimension $d$.  The recommendation is to select  larger values for data expansion than parameter expansion. In addition, using correlated indices require smaller values for these hyper-parameters. Figure \ref{fig:log.sim.10k.2} and \ref{fig:log.sim.10k.10} present the simulation results for the scenarios with $N=10,000$ and $d\in\{2,10\}$. In the supplemental material we include two additional scenarios  $N=1,000,d=2$ and $N=10,000,d=2$. The height of the bars represents the value of the relative measure and we add the dashed line at 1 to make it easier to separate performance improvements from deteriorations.

From Figures \ref{fig:log.sim.10k.2} and \ref{fig:log.sim.10k.10} a few lessons emerge. Combining PG with DAC produces good results, likely because  the Gaussian approximation is accurate  for such a large sample. The ESS/cpu grows with the number of batches.

The performance of the  coreset-based  algorithm yield a relatively high  RMSE. 
To shed some light on this performance we can recover from \cite{huggins2016coresets} the discrepancy, $\eps$, between the original likelihood and the coreset one,  as a function of the coreset size, mean sensitivity, and parameter dimension for $\delta=0.10$.

 Table~\ref{table:core.eps} shows the  average (over 50 replicates) discrepancy $\eps$ for different values of the sample size $N$, parameter dimension $d$, and data fraction $f$ divided by the average maximum value of the full data likelihood.
\begin{table} [!ht]
\begin{center}
\caption{ Coreset (logistic model): Relative average discrepancy $\eps$ for different values of sample size $N$, parameter dimension $d$, and data fraction  $f$. The numbers represent the average discrepancy divided by the average maximum value of the full data likelihood. }
\smallskip
\scalebox{1.0}{
\begin{tabular}{l || c c | c c  }
\multicolumn{1}{c}{ }  &  \multicolumn{2}{c}{$N=1,000$}   & \multicolumn{2}{c}{$N=10,000$}\\
\hline
Fraction         & $d=2$  &  $d=10$ & $d=2$  &  $d=10$  \\
\hline
$f=0.50$         &   3.217 & 3.602       &     0.685 & 1.160         \\
$f=0.10$         &   7.845 & 9.717       &     3.570 & 3.154           \\
$f=0.05$         &   11.992 & 13.972     &     6.138 & 4.461          \\
$f=0.01$         &   28.595 & 31.663     &     16.383 & 10.091         \\
\hline
\end{tabular}
}
\label{table:core.eps}
\end{center}
\end{table}
It is not surprising that $\eps$ increases as the fraction (i.e, the coreset size) decreases, but we also  can see that the discrepancy is generally quite large and this explains the poor performance of the sampler. The ESS/cpu measure beats PG only when using 1\% of the samples, but this comes at the expense of a vastly inflated RMSE.

Overall, subsampling techniques show good results with very high ESS/cpu without sacrificing the accuracy of the posterior, when $d=2$. The logistic posterior tends to be unimodal so the parameter expansion methodology is more suitable and clearly a larger concentration is achieved for $d=2$ than for $d=10$. The deterioration of the performance is clearly visible for $d=10$ although the method still controls the RMSE at the PG level. Since  the data do not exhibit any  clusters, it is not surprising that the data expansion techniques are not competitive to the parameter expansion ones.


Note that the computational time for the calculation of the log-likelihood can be significantly reduced using the vectorization trick available in R. This method allows much faster calculation by executing operations on the entire vectors of data instead of using a 'for' loop that goes through all the $N$ records one by one. This technique enabled us to increase the sample size to $100,000$. The comparison of the samplers using the vectorization-induced speed-up can be found in the Supporting Materials. 

\subsubsection{German Credit Data}
This concerns  data with a sample size that is not exceedingly large, but the dimension of the parameter is higher than we have considered so far. Specifically, the german credit data consist of $1,000$ records and $49$ predictors including the intercept (see \cite {biswas2019estimating} for more information). Most predictors are dummy variables taking only 0 and 1 values. The target/response is binary, with 70\% of them being cases, so the response variable is quite balanced. Logistic regression is implemented to  predict 
 $\Pr(Y=1)$ from the set of features.  Before fitting the  model, we transform all the quantitative features by subtracting the minimum value and dividing by the range so their values are in the $[0,1]$ interval. 

\begin{figure}[!ht]
\begin{center}
\includegraphics[scale=0.40]{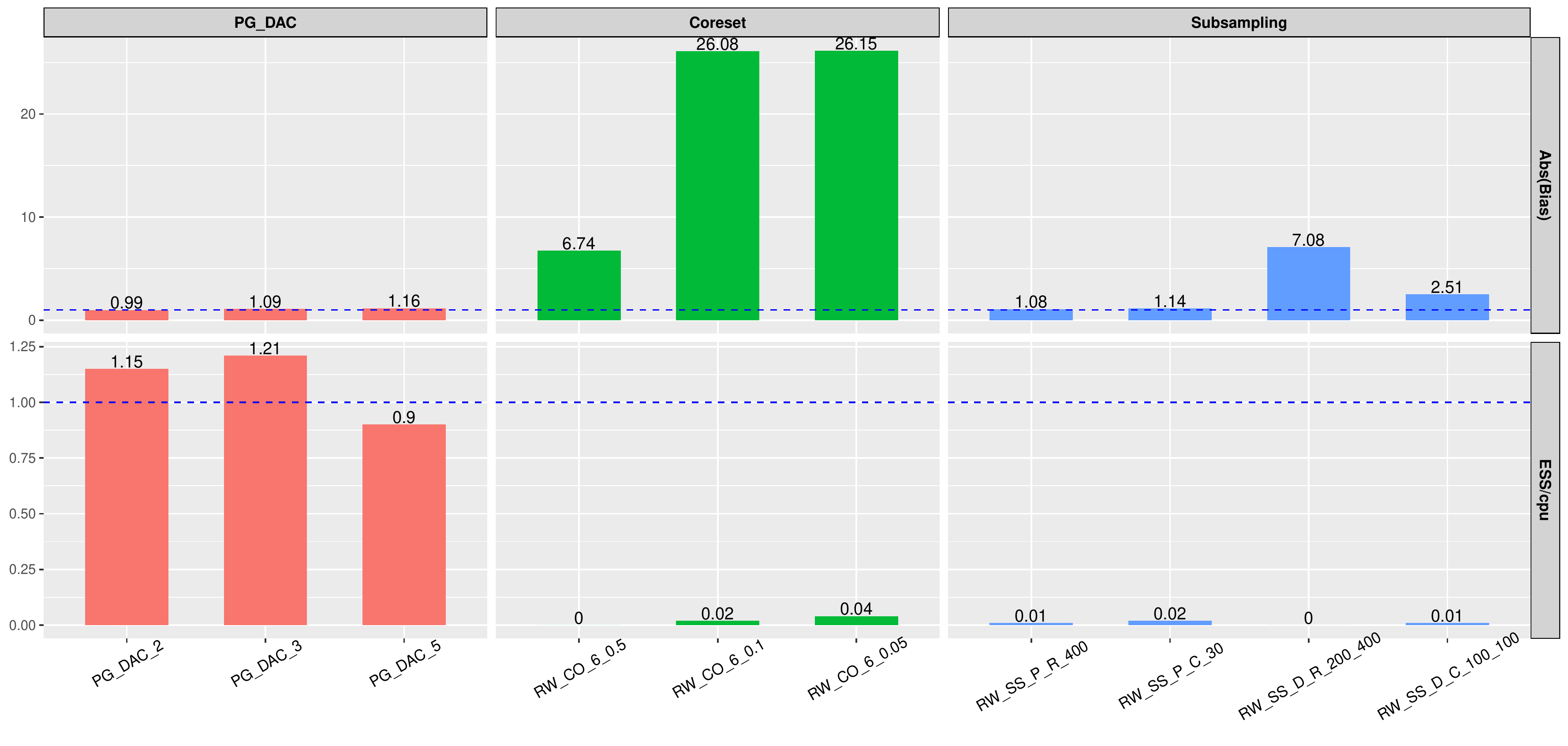}
\caption {German Credit data: Relative $\vert \mbox{Bias}\vert$ ({\em top row})and ESS/cpu ({\em bottom row})  for   DAC-based samplers ({\em left column}),  Coreset-based samplers ({\em center column}), and Subsampling-based samplers ({\em right column}).}
\label{fig:log.real}
\end{center}
\end{figure}

All the samplers are run for $N=100,000$ iterations,  burn-in is $B=50,000$ and adaptation occurs  every $500$ chain updates.  The performance of the samplers is presented in Figure~\ref{fig:log.real}. Note that the absolute value of the biases reported in the top row panels are calculated with respect to the maximum likelihood estimates,  as the true parameter values are not known. \color{black}We refer the reader to the Supplementary Materials for additional metrics and details.\color{black}

The results are similar to the ones obtained in the previous subsection, but some additional elements emerge. We can see that with 5 batches the DAC approach is losing a bit in terms of  bias  and even more on the ESS/cpu side. None of the subsampling-based methods (including coreset-based) can compete with the PG sampler, likely because the signal to noise ratio is altered too much when implementing any of these methods.  We should  also recognize that PG is a  Gibbs sampler which, unlike RWM samplers, will move at every iteration. This makes a bigger difference  when the parameter space has large  dimensions since then the RWM chain often gets stuck, especially if  the posterior exhibits strong dependence. 

%

Based on these numerical experiments, we conclude that with a very large sample size the first choice would be to use a DAC technique as long as the Gaussian approximation is likely to be accurate. The latter assessment will have to take into account the number of parameters and the nature of the model and data. If the Gaussian approximation is unsuitable, the subsampling methods can be used. The user will need to decide if the posterior is likely to be concentrated, so that they can use a parameter expansion, or the data exhibits multiple clusters, in which case a data expansion is needed. In the latter case, an exploratory analysis is recommended to determine reasonable values for the number of centroids, $K$.  The size of the subsample $m$ is typically decided based on the computational power available at the time of the analysis - we recommend using the largest possible value that can be handled by the system.

\subsection{Stochastic Volatility}
When analyzing stationary time series,  it is frequently observed that there are periods of high and low volatility, a phenomenon known as \textit{volatility clustering}  \citep[see for example][]{lux2000volatility}. One way to model such
behaviour is through a Stochastic Volatility (SV) model, where variances of the observed time series depend on hidden states that themselves form a stationary time series. We work with the following model which is indexed by parameter $\theta=(\bth_1,\bth_2,\bth_3)$:
\begin{equation}
\begin{split}
&x_{1} \sim \mathcal N(0,1/(1-\bth_1^2));\hsp v_i\overset{iid}{\sim}\mathcal N(0,1);\hsp w_i\overset{iid}{\sim}\mathcal N(0,1) ; \hsp i=\{1,\ldots,N\}, \\
&x_{i} = \bth_1 x_{i-1} + v_i; \hsp i=\{2,\ldots,N\}, \\
&y_{i} = \sqrt{\exp[\bth_2 + \exp(\bth_3)x_i]}w_i;  \hsp i=\{1,\ldots,N\}.
\end{split}
\end{equation}
Only data $\by=(y_1,\ldots,y_N)$ are observed, and $(x_1,\ldots,x_N)$ are latent/hidden states. The parameter $\bth_1\in (-1, 1)$  controls the auto-correlation of hidden states, while $\bth_2$ and $\bth_3$ are unrestricted and relate to the hidden states influence on the variability of the observed series. Given a hidden state, the distribution of the observed variable is Gaussian. We introduce the following priors, independently for each parameter:
\begin{equation}
\begin{split}
& \bth_1 \sim Unif[0,1], \\
& \bth_2 \sim \mathcal N(0,1), \\
& \bth_3 \sim \mathcal N(0,1).
\end{split}
\end{equation} 
We set the true parameters to $\bth_{true}=(0.95,-2,-1)$ and consider three lengths of the time series $N=100$, $500$ and $1,000$. Note that the model does not admit a closed form log-likelihood  but allows simulations of pseudo data sets. Therefore, for this model we only consider simulation-based ABC samplers:  AABC, BSL, ABSL, and the benchmark SMC ABC.
The summary statistics used for all the samplers is $S(\by)\in \RR^6$ and has the following components:
\begin{description}
\item[(C1)] Average of $\by^2$,
\item[(C2)] Standard deviation of $\by^2$,
\item[(C3)] Sum of the first 5 auto-correlations of $\by^2$,
\item[(C4)] Sum of the first 5 auto-correlations of binary series $\{\one_{\{y_i^2<\mbox{quantile}(\by^2,0.1)\}}\}_{i=1}^N$,
\item[(C5)] Sum of the first 5 auto-correlations of binary series $\{\one_{\{y_i^2<\mbox{quantile}(\by^2,0.5)\}}\}_{i=1}^N$,
\item[(C6)] Sum of the first 5 auto-correlations of binary series $\{\one_{\{y_i^2<\mbox{quantile}(\by^2,0.9)\}}\}_{i=1}^N$.
\end{description}
The $\mbox{quantile}(\by,\tau)$ is defined as the $\tau$-th quantile of the sequence 
$\by$. We focus here on $\by^2$ and its auto-correlations because the model parameters only affect its variability; the auto-correlation of $\by$ is zero for any lag. The  components (C4)-(C6) have been considered because the auto-correlations  of those binary series, defined under different quantiles, are useful in characterizing a time series \citep{schmitt2015quantile,dette2015copulas}.  The ABC, AABC, BSL and ABSL samplers are run for $M=55,000$ iterations. The burn-in period is of length $B=15,000$, with adaptation taking place every other $200$ iterations. 

\begin{figure}[!ht]
\begin{center}
\includegraphics[scale=0.40]{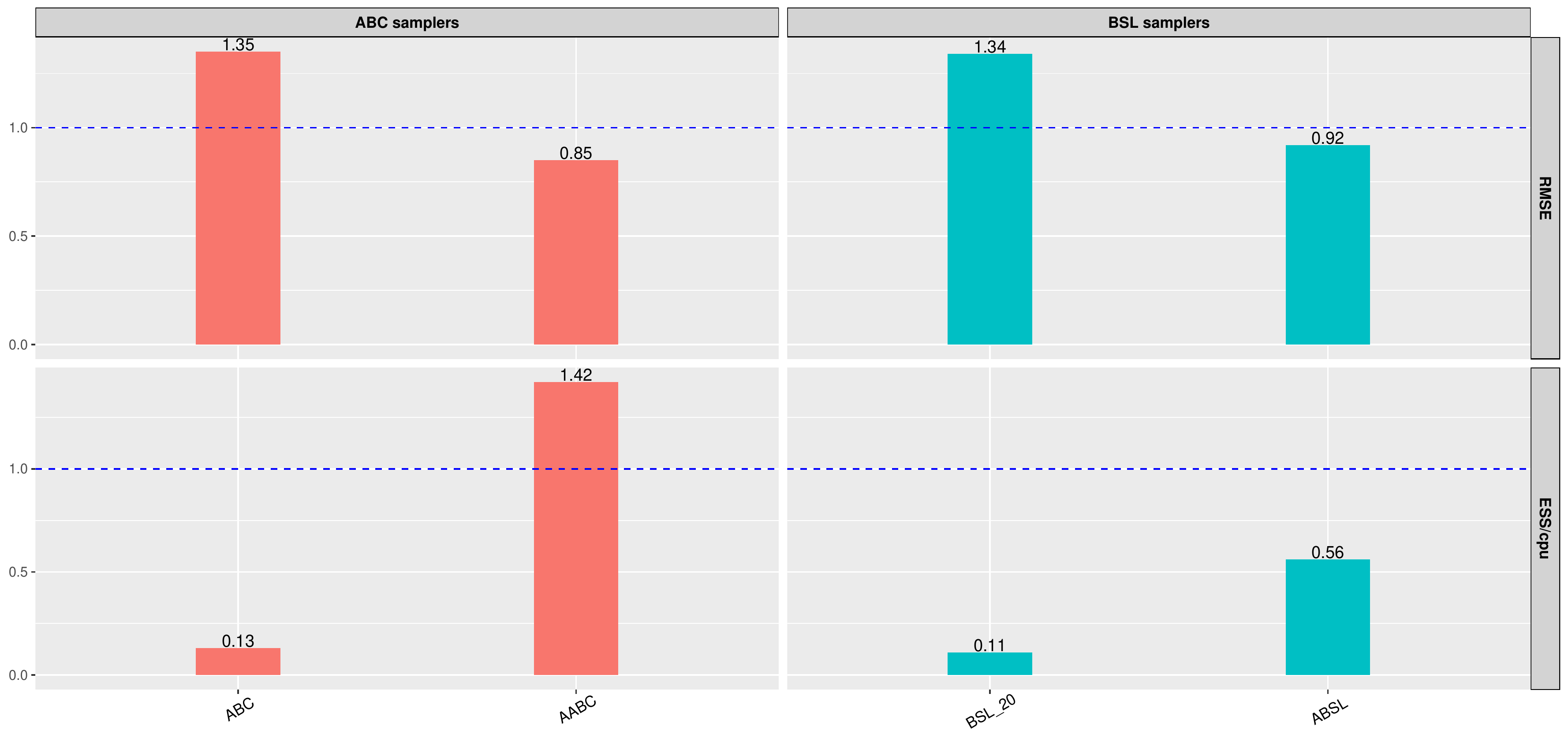}
\caption {SV model:  Relative RMSE ({\em top row}) and ESS/cpu ({\em bottom row}) when $N=500$ for ABC-based samplers ({\em left column}) and BSL-based samplers ({\em right column}).
}
\label{fig:sv.sim.500}
\end{center}
\end{figure}

\begin{figure}[!ht]
\begin{center}
\includegraphics[scale=0.40]{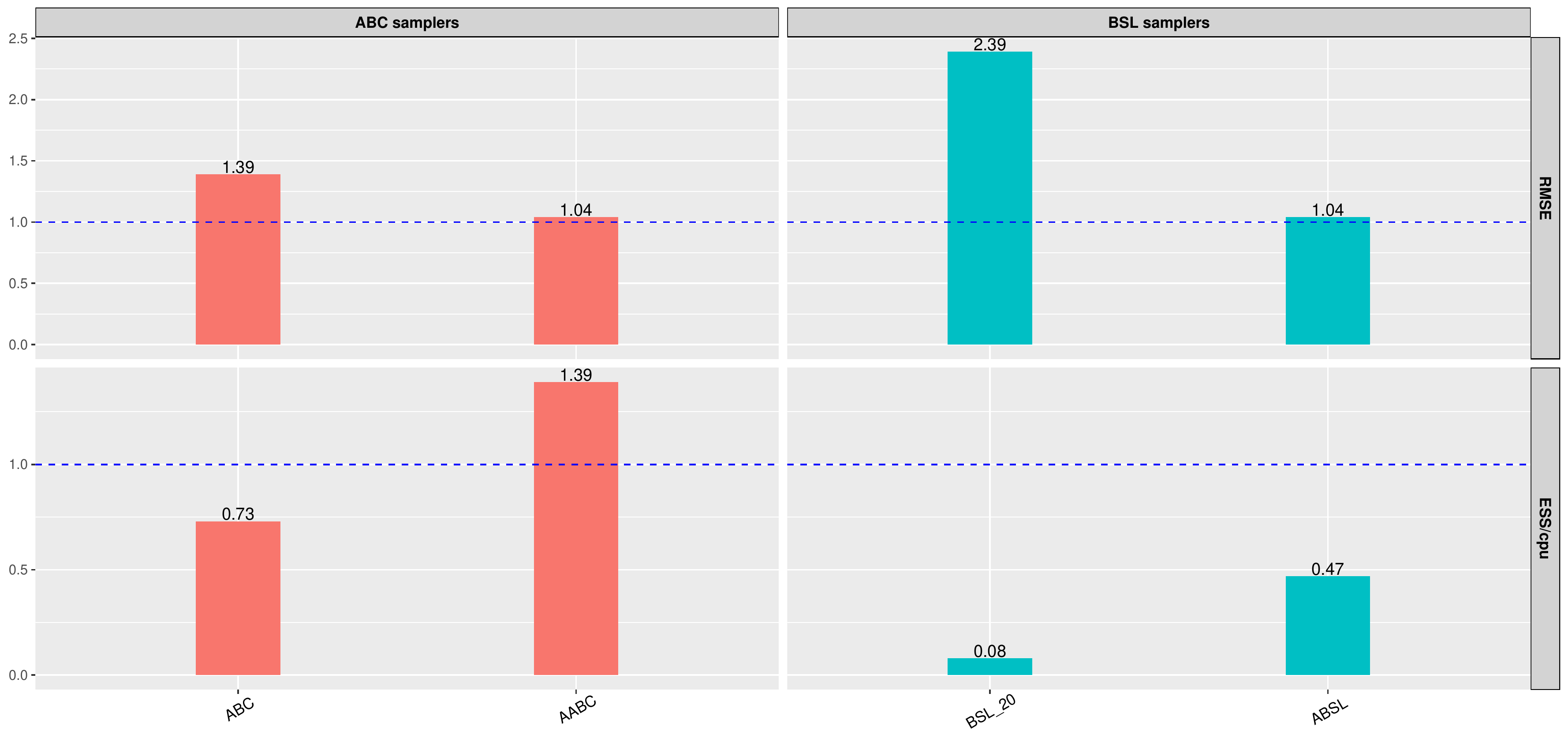}
\caption {SV model:  Relative RMSE ({\em top row}) and ESS/cpu ({\em bottom row}) when $N=1000$ for ABC-based samplers ({\em left column}) and BSL-based samplers ({\em right column}).}
\label{fig:sv.sim.1000}
\end{center}
\end{figure}

Figures~\ref{fig:sv.sim.500} and \ref{fig:sv.sim.1000} present the simulation results when  $N=500$ and, respectively, $N=1000$. The ABC and BSL samplers exhibit loss in terms of both RMSE and ESS/cpu when compared to the benchmark. The BSL is more costly since we generate 20 pseudo-data sets at each iteration. Not surprisingly, using pre-computation designs reduces the CPU time so we see a bump in efficiency for AABC and ABSL. Less obvious is the reduction in RMSE which is due to the increase the number of pseudo-data one can use while still saving computational time and the  higher acceptance rate. These findings mirror those of \cite{levi2021finding} and we refer the reader to that paper for more in-depth explanations. In this example ABC-based samplers outperform the BSL ones. The likely reason is that the Gaussian approximation on which the BSL relies is not accurate for this choice of the summary statistic, $S(\by)$. 

Overall, we find reasons for cautious optimism in these numerical results. They show that careful and controlled injection of  noise in the transition kernel can bring real practical benefits.

\section{Conclusion and future directions}

The Bayesian computational community finds itself at an inflexion point. Traditional MCMC computation is no longer tenable for complex problems. The new ideas and developments discussed here reduce significantly the computational costs  or bypass the intractability of the likelihood, but introduce additional  layers of approximation. The latter  requires a careful theoretical analysis to make sure that incurred errors are realistically controllable via tuning parameters. 

Complex models are often defined using high-dimensional parameters. MCMC methods  sample efficiently high-dimensional spaces as long as there are no bottlenecks or regions of small probability that the chain has difficulty traversing. Adaptive MCMC methods \citep{tut-amcmc,hoffman2014no, yang2019adaptive, pompe2020framework} have been proven effective for sampling in high-dimensional spaces with unfriendly geometries. Injecting adaptive ideas into the world of sampling with intractable targets is hindered by  stringent conditions that need to be satisfied by an adaptive transition kernel, e.g. the containment condition  \citep{bai2011containment,latuszynski2014containment}. Some inroads  have been made into eliminating the latter  in \cite{craiu2015stability} and \cite{rosenthal2018ergodicity} so we expect to see more adaptive designs permeating in pseudodata-generation-type samplers.

Constraints on paper length and considerations of subject matter consistency have prevented us from discussing methods that do not rely on MCMC sampling to perform Bayesian inference such as variational Bayes \citep{blei2017variational} or integrated nested Laplace approximation \citep{rue2017}. 
These are active research threads that continue to develop rapidly under the impetus provided by the expansion of data science, explosive growth of machine learning methods and other computationally demanding domains of information processing. Creative intertwining of most of the ideas or methods mentioned in this paper  will likely continue well into the future, but we believe that entirely new perspectives are also necessary in order to create the automatization of computation that is required if widespread use of Bayesian methods is to be seen in the 21st century. 
 
\section*{Acknowledgements} This research has been funded by NSERC of Canada. The authors thank Nancy Reid for the invitation to write this article, Alicia Carriquiri for guidance in defining the paper's scope, and an anonymous referee for a number of suggestions that have led to important improvements.

\bibliographystyle{ar-style1}
\bibliography{superref}

\end{document}